\documentclass[conference]{IEEEtran}
\IEEEoverridecommandlockouts

\usepackage{cite}
\usepackage{amsmath,amssymb,amsfonts}
\usepackage[hidelinks,urlcolor=blue]{hyperref} 
\usepackage{graphicx}
\usepackage{textcomp}
\usepackage{xcolor}
\usepackage{gensymb}
\usepackage{algorithm}
\usepackage{algorithmic}
\graphicspath{{figures/}}
\def\BibTeX{{\rm B\kern-.05em{\sc i\kern-.025em b}\kern-.08em
    T\kern-.1667em\lower.7ex\hbox{E}\kern-.125emX}}

\begin{document}

\title{Gate Voltage Effect on Pulse Detection Efficiency of Perimeter-Gated SPADs

\thanks{This material is based on work supported by the National Science Foundation under Grant No. 2442346. Any opinions, findings, and conclusions or recommendations expressed in this material are those of the authors and do not necessarily reflect the views of the National Science Foundation.}
}

\author{\IEEEauthorblockN{Hunter Guthrie, Md Sakibur Sajal, Zexi Liu, and Marc Dandin}
{Department of Electrical and Computer Engineering},\\
{Carnegie Mellon University,}
{Pittsburgh, Pennsylvania, 15213, USA}\\
\\
{email: mdandin@andrew.cmu.edu}   }


\maketitle

\begin{abstract}
Perimeter-gated single-photon avalanche diodes (pg-SPADs) are known for their dynamic dark noise modulation capabilities. They are reported to trade noise for photon sensitivity under continuous illumination. However, the implications of this trade-off have not heretofore been studied with pulsed optical systems. This work bridges this gap. We demonstrate that pg-SPADs fabricated in a $\mathbf{0.35~\mu m}$ standard CMOS process trade-off pulse detection efficiency for a reduction in the the spread of spurious events within a burst window. Consequently, herein, we propose guidelines for the optimal use of pg-SPADs in pulsed LIDAR applications in view of the observed trade-off. 
\end{abstract}


\section{Introduction}
Single-photon avalanche diodes (SPADs) are a widely used technology in pulsed optical sensing systems such as LIDAR~\cite{pulsedlidar1,pulsedlidar12,pulsedlidar13,pulsedlidar14,pulsedlidar15}, fluorescence sensing~\cite{spadfl1,spadfl2,spadfl3}, and time-resolved imaging~\cite{spadtime1,spadtime2,spadtime3}, owing to their fast timing response. Specifically, in advance LIDAR systems, the practice is being shifted from single pulse-mode to burst-mode where a modulation or a code is used to send multiple closely spaced pulses in the air~\cite{Kang:24}. For these applications, the registration of photon arrivals are correlated with the short burst windows, which is critical for measurement fidelity. However, the core limitation of SPAD-based detectors is the occurrence of spurious detections, including thermally generated dark carriers independent of photon arrivals~\cite{Xu2017ComprehensiveTechnologies,dcr2,dcr3,dcr4,dcr5} and afterpulsing due to the trapped charges from previous avalanche events~\cite{apl1,apl2,apl3}. 
\begin{figure}
    \centering
    \includegraphics[width=\linewidth]{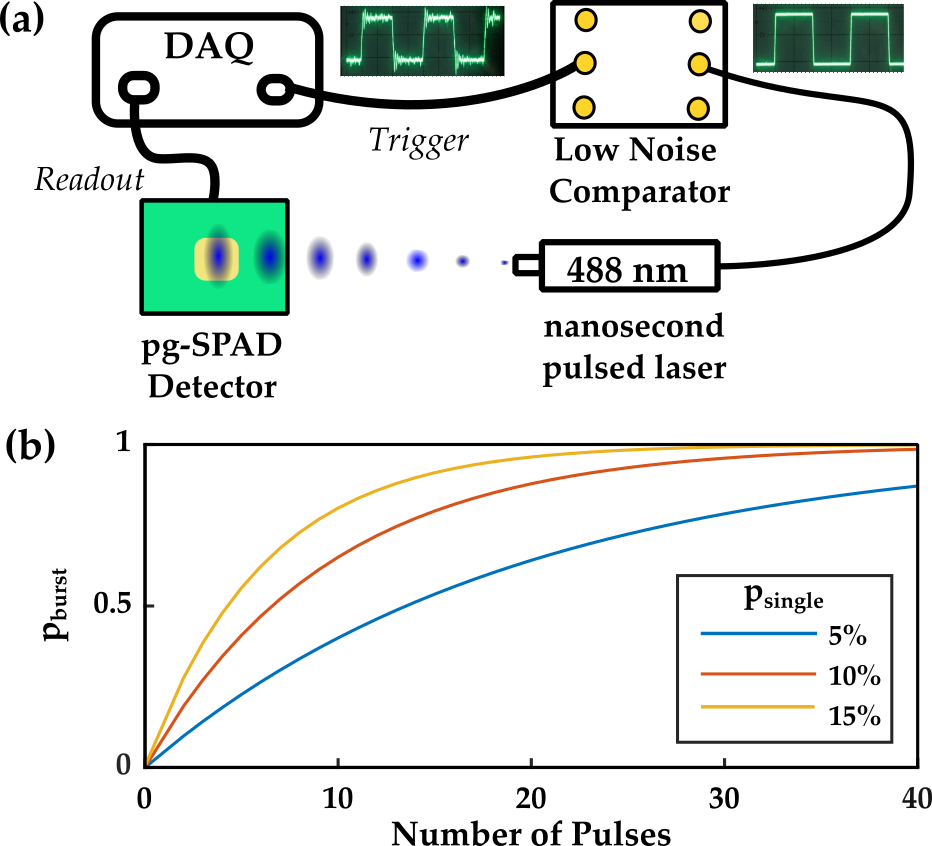}
    \vspace{-10pt}
    \caption{(a) SPAD based pulse detection system using a perimeter-gated SPAD (pg-SPAD) detector and (b) the probability of burst detection as a function of the number of pulses in a burst with different pulse detection rate.}
    \vspace{-15pt}
    \label{fig:pulse}
\end{figure}

For example, Figure~\ref{fig:pulse} (a) shows our system diagram where a data acquisition module (DAQ) sends a pulse train to trigger a $488~nm$ nanosecond pulsed laser. However, since the dark carriers are generated at all times in addition to the photon incident window, they reduce the correlation between the input pulse train and the recorded signal. 
\begin{figure*}[h]
    \centering
    \includegraphics[width=\linewidth]{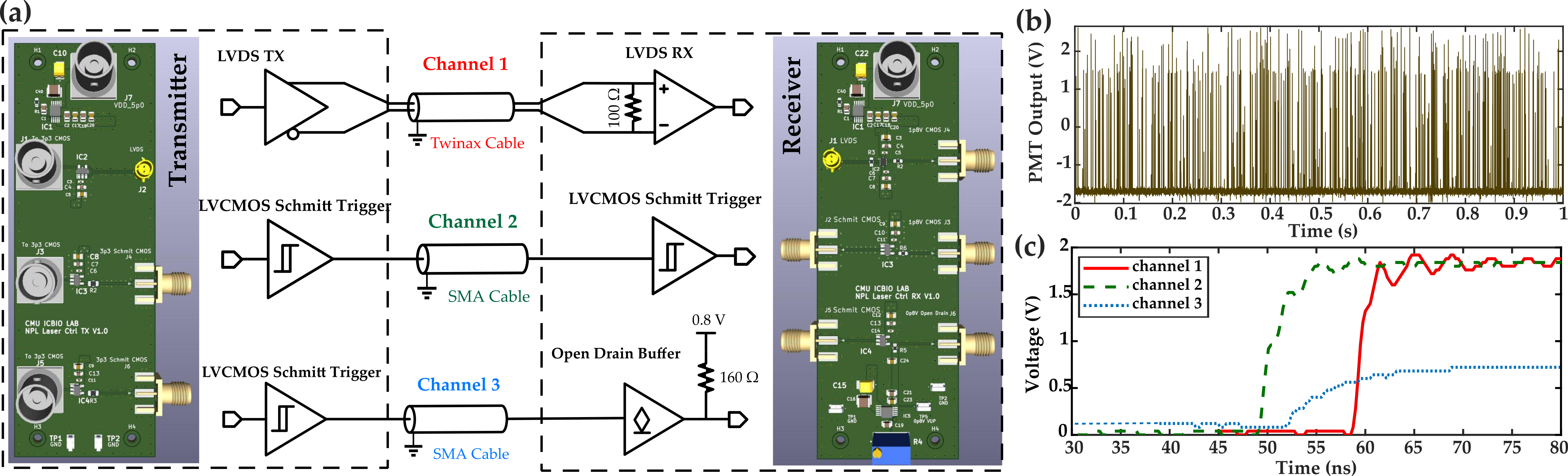}
      \vspace{-10pt}
    \caption{(a) Transmitter and receiver PCB for relying a noise-free trigger signal from the DAQ to the laser. (b) False triggering caused by noise when the denoising PCB is bypassed. Laser output was measured using a photomultiplier tube (PMT). (c) Outputs of the three channels from the denoising PCB.}
      \vspace{-10pt}
    \label{fig:background}
\end{figure*}

While there are several structural techniques for suppressing the dark noise~\cite{Finkelstein2006STI-BoundedTechnology, Richardson2011ScaleableTechnology}, perimeter gating offers a simple and technology-agnostic mechanism to do the same~\cite{Dandin2010, Nouri2012,  Dandin2012a, Dandin2016, Dandin2017a,sajalpbit,nicoleHabib}. It consists of placing a poly-silicon gate at the diode's periphery in order to actively reduce the strength of the peripheral electric field and prevent premature edge breakdown. These devices are known as perimeter-gated SPADs (pg-SPADs).

Recent works~\cite{10405999,10115088} have investigated the benefits of pg-SPADs through simulating noise suppression in LIDAR applications.  Although the perimeter gating technique has been found to trade noise with sensitivity to photons under constant illumination~\cite{Dandin2012a}, the impact of it on pulse-detection under realistic excitation conditions remains unexplored. For example, assuming that the probability of a single pulse detection is $p_{single}$, the cumulative probability of detecting at least one pulse in a burst of $N$ pulses becomes
\begin{equation}
\label{eq:burst_prob}
    p_{burst} = 1 - (1-p_{single})^{N}.
\end{equation}
Figure~\ref{fig:pulse} (b) shows $p_{burst}$ as a function of $N$ for three rates of $p_{single}$, \textit{i.e.}, $5\%$, $10\%$, and $15\%$. We see that the number of pulses required in a burst for a guaranteed burst detection increases as the value of $p_{single}$ decreases. Hence, perimeter gating could reduce burst detection efficiency as a side effect of noise reduction. However, a cost-benefit analysis is needed for a better understanding of its impact on the overall gain.

To that end, this work explored the effect of perimeter gate voltage on $p_{burst}$ as well as $p_{single}$ using a $64\times64$ array of custom pg-SPADs. These devices were fabricated in a $0.35~\mu m$ standard CMOS process with an integrated digital counter. We quantified the detection activity before, during, and after the optical excitation window (burst window) for a range of applied gate voltages and excess bias voltages. Our investigation revealed that both pulse and burst detection efficiencies decrease as a side effect of noise suppression. However, the spreads of spurious detections stemming from dark carriers and afterpulses within the burst windows are also found to be greatly reduced. As such, we introduced a practical guideline to optimally use pg-SPADs as detectors in a pulsed optical system by balancing the aforementioned trade-offs.

\section{Experimental Setup}
To investigate the effect of perimeter gate voltage on $p_{single}$ as well as $p_{burst}$, we constructed the setup shown in Fig.~\ref{fig:pulse} (a). A data acquisition (DAQ) system (NI-PXIe 8821 controller with NI-PXIe 6544 digital waveform generator, National Instruments) was used to trigger a 488 $nm$ commercial nanosecond pulsed laser system with an adjustable pulse width of 6-39 ns (NLP49B, Thorlabs). The pg-SPAD array was placed close to the laser to improve photon collection efficiency since the pixels have a $6\%$ fill-factor and no anti-reflection coating or lenses to compensate for that. A detailed description of the chip and its operation can be found in Refs.~\cite{Sajal2022Perimeter-GatedProbability,spadpuf}. For completeness, we briefly describe its general operation here.

Any pixel on the $64 \times 64$ array can be selected independently and multiplexed to the 12-bit on-chip output counter. The selected pg-SPAD is precharged beyond its breakdown voltage while the counter is reset at the same time before a measurement session starts. Once the device enters the integration phase, active quench and active reset signal quenches the avalanche current upon detection of a breakdown event and subsequently recharges the device to detect the next avalanche. The output of the selected pg-SPAD continuously increments the counter which is monitored by the DAQ at $F_s =$ 100~MHz to produce a time-series data of the detected events. 

\section{Experimental Methods}
\subsection{Pulse Generation}
Laser excitation with a fixed pulse width of $39~ns$ was used in this study. While a precise number of pulses can be sent using the DAQ to trigger the laser accordingly, the noise sensitive input port of the trigger was exhibiting auto-triggering in addition to the intended pulses. Measurements indicate that noise transients as small as $20~mV$ are sufficient to induce false triggering (see Fig.~\ref{fig:background} (b)), thereby degrading the fidelity of the measurements. Since this is critical for this study to ensure the timing and the number of transmitted laser pulses, dedicated circuits (see Fig.~\ref{fig:background} (a)) employing low-voltage differential signaling (LVDS), low-voltage CMOS (LVCMOS) and open-drain protocols are developed to transmit the laser activation signal with improved noise immunity.

The first channel incorporates a differential line driver that converts the single-ended LVCMOS output of the DAQ into LVDS signals. The differential signals are transmitted over a $100~\Omega$ twisted pair cable and converted back to be single-ended right before entering the laser trigger input. For LVCMOS and open drain channels, the laser activation signal is first fed into a Schmitt trigger. By introducing an input hysteresis of about $0.7~V$, the Schmitt trigger suppresses false triggering caused by noise or slow input crossings. Both channels leverage a SMA cable to minimize loop area and hence the EMI coupling. All components are powered by a low dropout regulator with a power supply rejection ratio greater than 70~dB to ensure power supply noise is minimized. We have designed these generalized boards to fit different system requirements (see Fig.~\ref{fig:background} (c) for different channel output profiles). The design files are available through IEEE DataPort~\cite{zexiBoard}.

\subsection{Detection}
During each measurement, the counter output continuously increases when an avalanche event takes place within the active period ($T_{active} = 4~ms$). During this time, a burst of $N = 10\sim200$ pulses was sent unless dark measurements were recorded. To enable averaging, one thousand measurements were made for each setting/condition. Custom MATLAB routines were used for sweeping, recording and analyzing the data. Using the pulse train signal, the data were segmented into three regions:
\begin{itemize}
    \item \textit{Pre-burst} window: prior to laser pulses for dark counts
    \item \textit{Burst} window: for the signal and the background
    \item \textit{Post-burst} window: for afterpulses and dark counts.
\end{itemize}

The \textit{peak} detection response was defined as the maximum incremental count within the entire window, referenced from the activation of the detector. Measurements were repeated across a range of perimeter gate voltage, $V_G = 0\sim4~V$ and excess bias voltage, $V_{ex} = 3\sim5~V$. The underlying assumption is that $V_{ex}$ and $VG$ function as tuning knobs to modulate pulse detection probability, $p_{single}$ while trading with the dark count rate (DCR). Here, the increase of $V_{ex}$ and $VG$ respectively increase and decrease both $p_{single}$ and DCR.

\begin{figure}
    \centering
    \includegraphics[width=\linewidth]{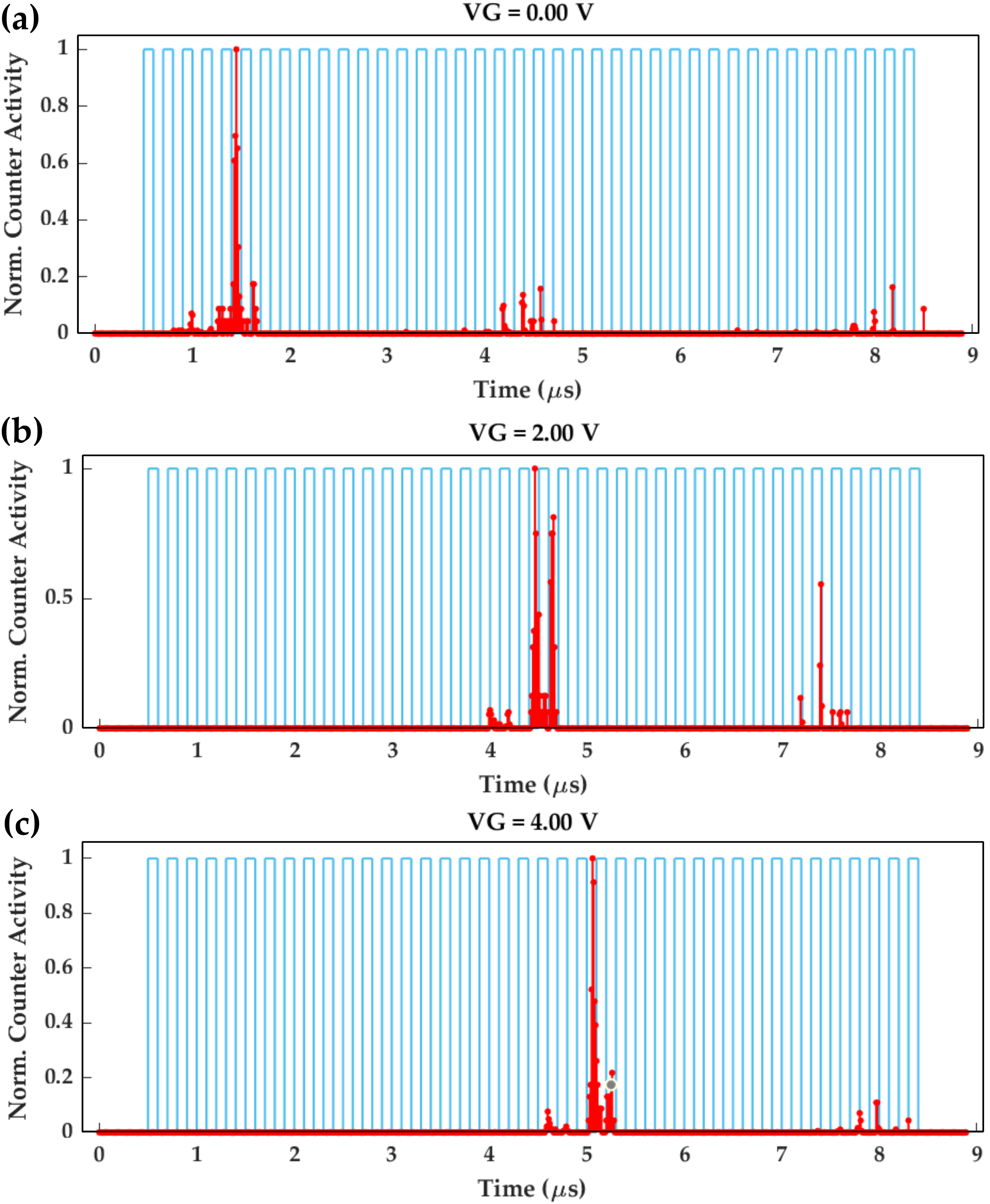}
    \vspace{-15pt}
    \caption{Time series data of the counter for different gate voltages, \textit{i.e.}, $VG = 0,2,$ and $4~V$ overlaid with the laser trigger signal within the burst window.}
    \vspace{-15pt}
    \label{fig:timeSeries}
\end{figure}

\subsection{Dark Count and Afterpulsing Analysis}
Separate measurements with the laser disabled were used to determine the DCR as a function of $V_{ex}$ and $VG$. These data were used to normalize signal-to-background metrics and isolate illumination-induced detections. 

Afterpulsing was quantified as the excess detection activity in the post-burst window \textit{relative} to the pre-burst DCR baseline. This provided a measure of the correlated noise following the laser excitation.

\section{Experimental Observations}

\subsection{Reduction of Photon Sensitivity}
Figures~\ref{fig:timeSeries} (a-c) show instances of normalized time series data of the counter increments during the burst window as $V_G$ is varied from $0\sim4~V$ with $V_{ex}$ set to $4~V$ and the number of pulses set to $40$. We can see that the activity peak shifts to the right as we increase the gate voltage magnitude. In other words, it takes more pulses to incident on the pg-SPAD for the burst detection. This observation agrees with the assumption that increased gate voltage reduces photon sensitivity as the junction electric field originated from the reverse bias voltage is modulated by the perimeter gate induced electric field.

To study the burst detection efficiency ($\eta_{burst}$) of the pg-SPAD, the number of pulses was varied from 10 to 200 pulses while varying the gate voltage from $V_G=0\sim4~V$ as well. The noise activity ($N_{noise}$) was determined by averaging the pre- and post-burst activity. The burst activity ($N_{burst}$) was determined by subtracting $N_{noise}$ from the counter activity within the burst window ($N_{total}$). Finally, we derived $\eta_{burst}$ by dividing $N_{burst}$ by the total number of pulses ($N_{pulse}$) sent. Mathematically,

\begin{equation}
    \eta_{burst} = \frac{N_{burst}}{N_{pulse}} = \frac{N_{total}-N_{noise}}{N_{pulse}}.
\end{equation}

\begin{figure}
    \centering
    
    \includegraphics[width=\linewidth]{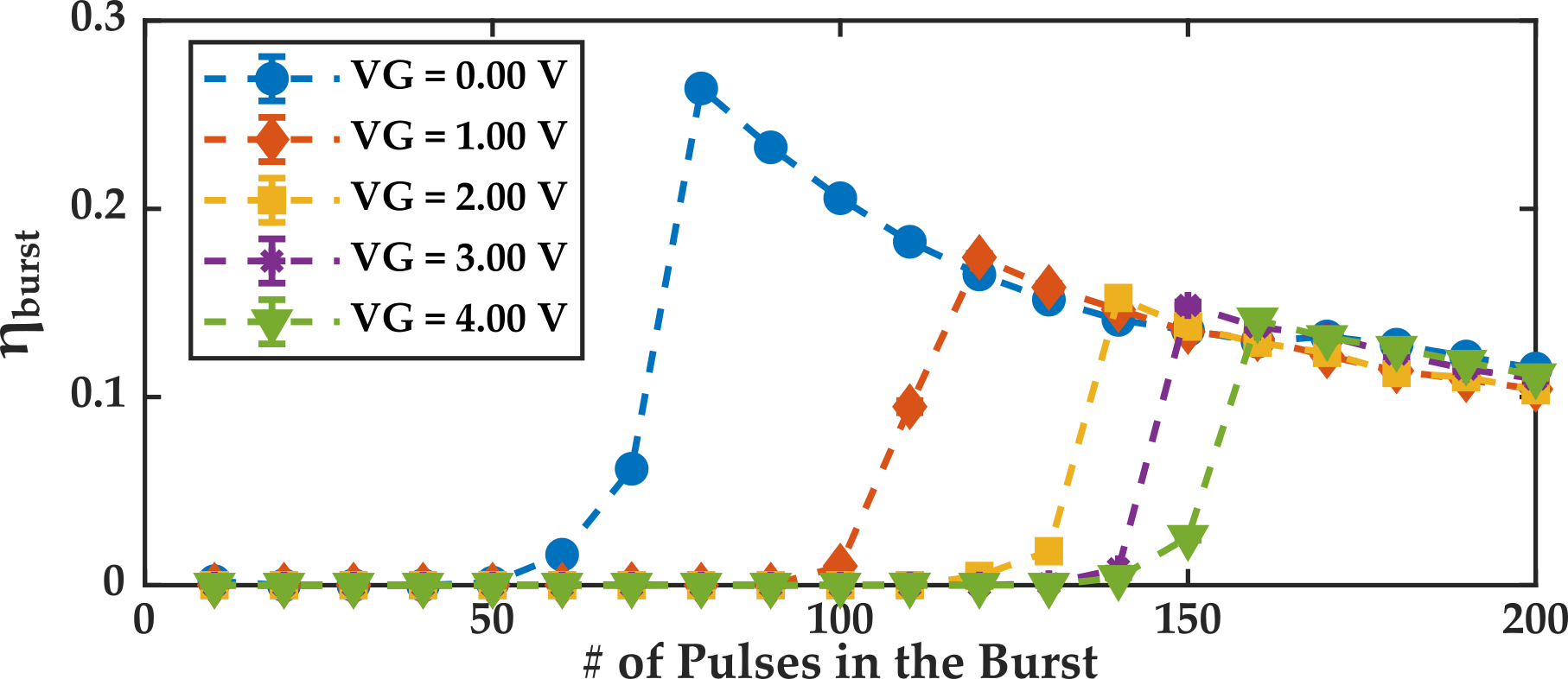}
    \vspace{-15pt}
    \caption{Burst detection efficiency at $V_{ex} = 4~V$ for $N = 10\sim200$.}
    \label{fig:eta}
\end{figure}

\begin{figure}
    \centering
    
    \includegraphics[width=\linewidth]{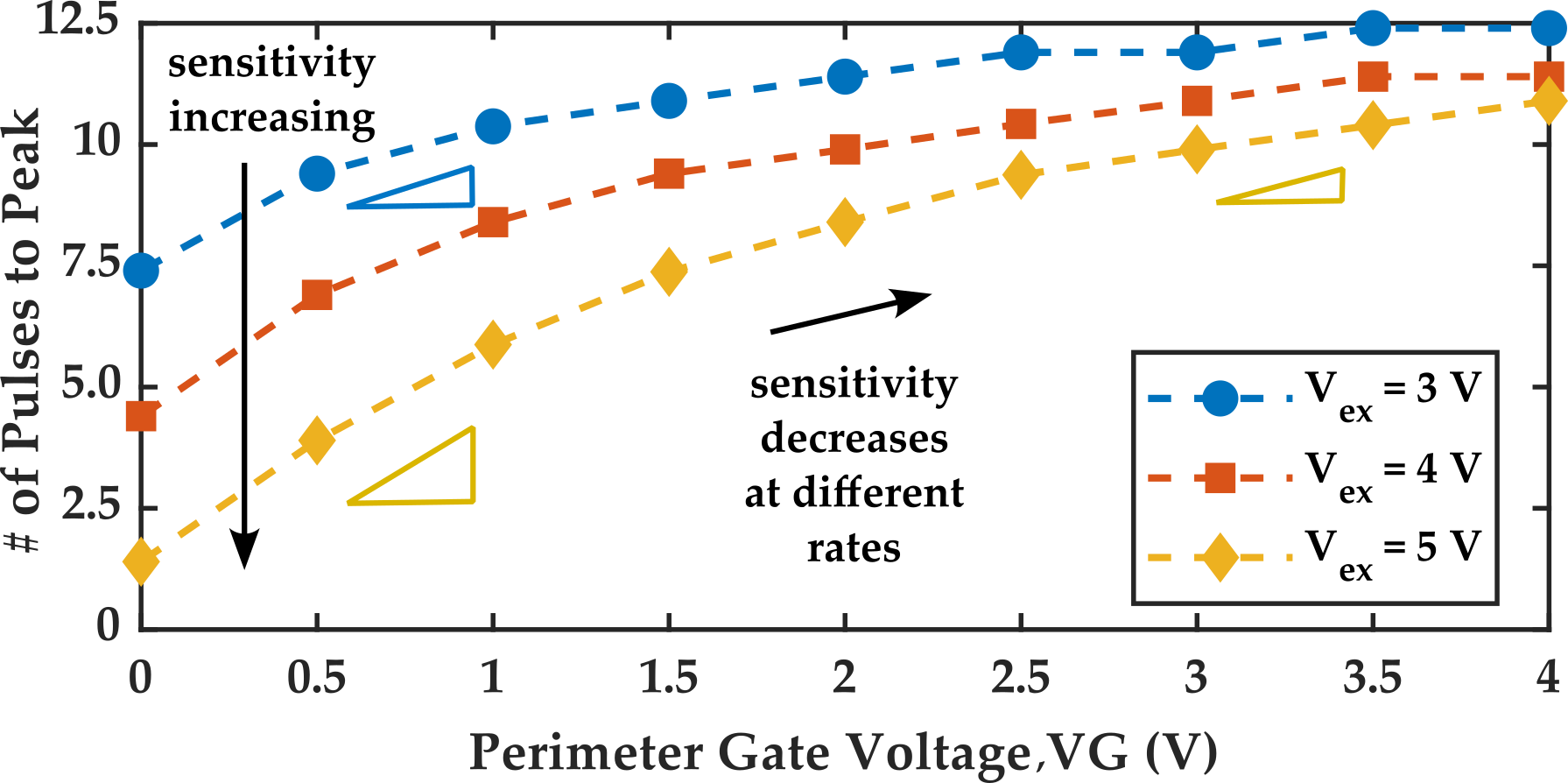}
    \vspace{-15pt}
    \caption{Minimum number of pulses to activity peak as a function of gate voltage and $V_{ex}$. Here, the slope represents the sensitivity on the gate voltage.}
    \vspace{-15pt}
    \label{fig:eta}
\end{figure}

\begin{figure*}[h]
    \centering
    \includegraphics[width=\linewidth]{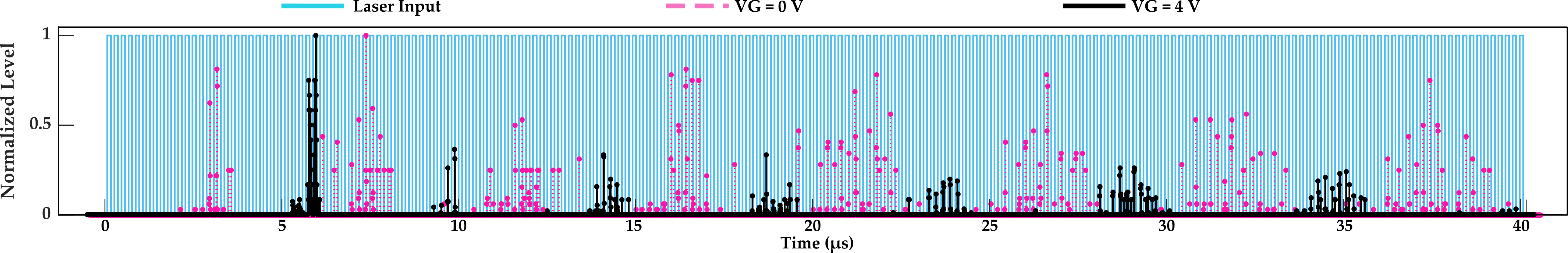}
    \vspace{-15pt}
    \caption{Long burst window showing the benefit of perimeter gate voltage in dark noise and afterpulsing reduction with $V_{ex} = 4~V$ and 200 pulses.}
    \vspace{-15pt}
    \label{fig:longburst}
\end{figure*}

Figure~\ref{fig:eta} (a) shows $\eta_{burst}$ as a function of the number of pulses in a burst and $V_G$. From this, it is clear that increase of gate voltage magnitude decreases the pulse detection efficiency. This is a direct trade-off of noise reduction unless the avalanche probability is boosted by increasing the excess bias voltage ($V_{ex}$) as discussed in the next subsection.

\subsection{Increase of Photon Sensitivity}

\begin{figure}
    \centering
    \includegraphics[width=\linewidth]{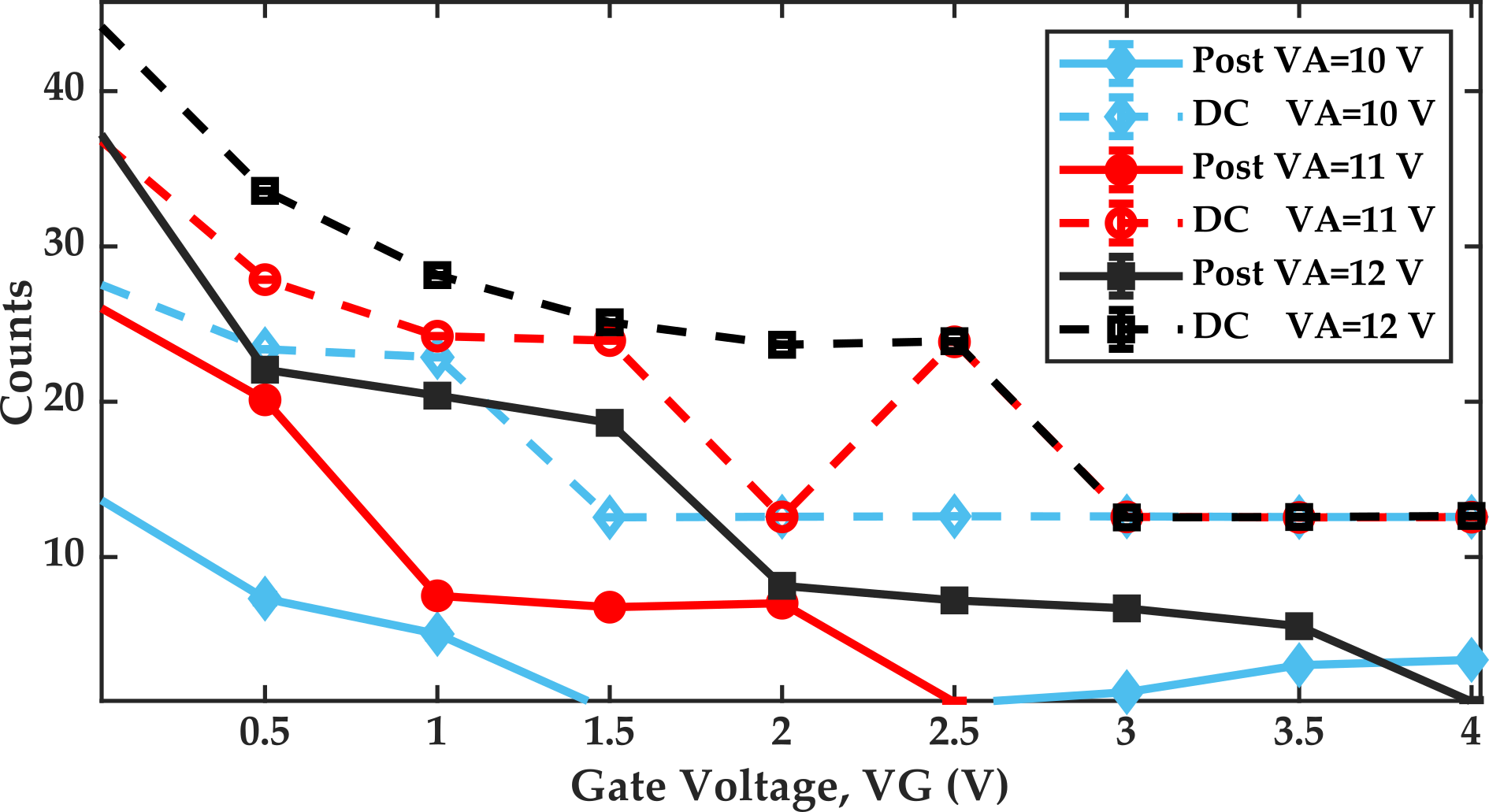}
    \vspace{-15pt}
    \caption{Post-burst activity in reference to DCR at different excess bias voltage.}
    \vspace{-15pt}
    \label{fig:afterpulse}
\end{figure}

Figure~\ref{fig:eta} (b) shows the effect of increasing the avalanche probability as well as $p_{single}$ through the increase of $V_{ex}$. As a result, the number of pulses required to achieve peak activity decreases. This was also predicted from Fig.~\ref{fig:pulse} (b) and Eq.~\ref{eq:burst_prob}. However, one interesting observation is that as we increase $VG$, the number of required pulses to peak activity gradually increases at different rates based on the excess bias voltage. Specifically, gate voltage sensitivity is higher for a higher $V_{ex}$. 

\subsection{Reduction in Spurious Detections}
Although increased $V_G$ reduced $\eta_{burst}$, the benefit of perimeter gating can be seen in long burst windows which are commonly used for long range LIDARs. Figure~\ref{fig:longburst} shows the time series data of a $40~\mu s$ long burst of $200$ pulses. An excess bias voltage of $4~V$ was used with $V_G = 0~V$ and $4~V$ to demonstrate the behavior of a native device and a pg-SPAD, respectively. We make two interesting observations. 

First, the native device exhibits detections spread out in wide clusters throughout the burst window. Second, the normalized activity level of all the clusters are similar, implying uniformly distributed dark noise and afterpulsing activity. However, for the pg-SPAD, we see a dominant first cluster despite occurring at a delayed time due to the reduction in the pulse detection probability. Furthermore, the spread of the remaining clusters is tighter compared to that of the native SPAD. Hence, we draw the conclusion that perimeter gate voltage can focus the avalanche activity within narrow time bins by suppressing the spurious activities for a given incoming photon rate.

Figure~\ref{fig:afterpulse} corroborates the above statement by analyzing the post-burst window with and without laser excitation. The broken lines show how the dark counts caused by the thermally generated carriers and the afterpulses vary with respect to $V_G$.

\section{Guidelines for using pg-SPADs}
In this section, we combine the findings of the present study with those from our previously reported work~\cite{Sajal2022Perimeter-GatedProbability}. Specifically, we propose a general guideline on how to optimally use pg-SPADs for pulsed optical systems. We have shown previously~\cite{Sajal2022Perimeter-GatedProbability} that the probability of dark counts can be reduced, if not eliminated, by using higher perimeter gate voltage. However, the magnitude of the gate voltage required to cause vanishing dark counts depends on the activation period and the excess bias voltage. Hence, for applications requiring longer activation times \textit{i.e.}, long range LIDARs or higher excess bias voltage, \textit{i.e.}, fluorescent lifetime imaging, higher gate voltage is preferred to suppress the dark noise.

However, this study showed that higher gate voltages reduce the pulse detection efficiency. As such, the gate voltage should be optimized based on the activation time and the excess bias voltage so that the SNR is maximized instead of overdriving the gate~\cite{Dandin2012a}. For a given activation time, the number of pulses can be optimized based on the $\eta_{burst}$ plots for the minimum gate voltage that achieves zero DCR~\cite{Sajal2022Perimeter-GatedProbability}. However, it should be noted that based on the strength of the returned signal, the number of pulses, and the bias voltages \textit{i.e.}, gate voltage and/or excess bias voltage need to be adjusted for optimal operation. 

Our future work will entail deploying a correlation algorithm for evaluating the effect of gate voltage on the ability to correlate N-pulse trains with the first detected peak, often used in pulsed time of flight systems, as well as the ability to correlate long frequency-modulated pulse trains. Additionally, a framework will be developed to actively adjust the gate voltage, reverse bias, and pulse train in order to leverage the aforementioned tradeoffs between pulse detection efficiency, DCR, and afterpulsing.

\section{Conclusions}
In this study we have explored the effect of perimeter gating on the pulse detection efficiency of pg-SPAD detectors equipped with a high-speed laser triggering system. Perimeter gating increases the number of laser pulses required to efficiently detect a pulse train of length N, indicating a degradation of pulse detection efficiency. However, we find that this may not be an issue for modern pulsed detection systems which attempt to correlate a well-defined pulse train with a return signal. While the registration efficiency for the first peak is degraded and shifted later in time, the afterpulsing and DCR are greatly reduced, leading to a more defined return signal with less noise and jitter. 

\newpage

\bibliographystyle{IEEEtranDOI}
\bibliography{main}


\end{document}